\documentstyle[psfig]{mn}

\newif\ifAMStwofonts
\AMStwofontstrue

\def\cm{\,{\rm cm}}
\def\erg{\,{\rm erg}}
\def\Hz{\,{\rm Hz}}
\def\sr{\,{\rm sr}}
\def\s{\,{\rm s}}
\def\K{\,{\rm K}}
\def\kms{\mbox{km\,s$^{-1}$}}

\def\expec#1{\langle#1\rangle}
\def\dd{\,{\rm d}}
\newcommand{\op}{Ly$\alpha$\ }
\newcommand{\hi}{\mbox{H{\scriptsize I}}}
\newcommand{\hei}{\mbox{He{\scriptsize I}}}
\newcommand{\heii}{\mbox{He{\scriptsize II}}}

\title[Probing the thermal history of the IGM]{Probing the thermal
history of the Intergalactic Medium with \op absorption lines}

\author[Martin G. Haehnelt and Matthias Steinmetz]
{Martin G. Haehnelt$^{1}$ and Matthias Steinmetz$^{1,2}$\\ 
${1}$ Max-Planck-Institut f\"ur Astrophysik, Postfach 1523, 85740
Garching, Germany \\
$^{2}$ Steward Observatory, University of Arizona, Tucson, AZ 85721, USA}

\pagerange{\pageref{firstpage}--\pageref{lastpage}}
\pubyear{1994}

\begin{document}

\maketitle

\label{firstpage}

\begin{abstract}
The Doppler parameter distribution of \op absorption is
calculated for a set of different reionization histories. 
The differences in  temperature between different reionization
histories are as large as a factor three to four depending on the 
spectrum of the ionizing sources and the redshift of helium reionization. 
These temperature differences result in observable differences in
the Doppler parameter distribution. Best agreement with the observed
Doppler parameter distribution between redshift two and four 
is found if hydrogen and helium are reionized simultaneously at or 
before redshift five  with a quasar-like spectrum.
\end{abstract}


\section[]{Introduction}

The physical conditions of the Intergalactic Medium (IGM) have been 
under discussion ever since its existence was postulated. 
The absence of strong  \op absorption in the spectra
of high-redshift objects showed that any homogeneous IGM  must be 
either highly ionized or extremely tenuous (Gunn \& Peterson 1965,
Bahcall \& Salpeter 1965). Nevertheless the IGM density and
temperature have been poorly constrained for many years. Both a hot 
collisionally ionized IGM heated by energy input
due to supernovae and a warm intergalactic medium heated by
photoionization  have been discussed extensively in the literature. 
Recent progress in numerical simulations including 
the relevant gas physics suggests that the low column density 
\op forest is due mainly to density fluctuations of moderate 
amplitude in  a warm photoionized phase of the IGM 
(Cen et al.~1994; Petitjean, M\"ucket \& Kates 1995; 
Zhang, Anninos \& Norman 1995; Hernquist et al.~1996; 
Miralda-Escud\'e et al.~1996; but see also  
Bi, B\"orner \& Chu 1992 for early analytical predictions).  
The baryon content in this warm phase must be similar  
to that predicted by cosmic nucleosynthesis to produce the observed 
mean absorption in the spectra of high-redshift quasars. 
At redshifts larger than two most of all  baryons are therefore  
probably in such a warm phase of the IGM and 
little room is left for an additional hot phase. 
(Rauch \& Haehnelt 1995, Rauch et al. 1997, Weinberg et al. 1997).
As pointed out by a number of authors the recombination time 
scale of the IGM becomes longer  than the Hubble time at a redshift 
of about five. As a result, the temperature of the warm IGM has  
some memory for when and how it was (re-)ionized 
(Miralda-Escud\'e \& Rees 1994, Meiksin 1994,  Hui \& Gnedin 1997).  
We  have used a set of hydrodynamical SPH simulations for different 
plausible reionization histories  to demonstrate that these do 
indeed result in  significant and observable differences in the thermal
history of the IGM.   In this letter  we investigate 
the observable consequences on the Doppler parameter distribution 
of \op absorption lines and speculate on the possibility to
discriminate between different reionization histories.

\begin{figure*}
\vskip -1.75cm
%\centerline{
%/for/noneq/solver/fig101.ps
\hskip-27.0cm\psfig{file=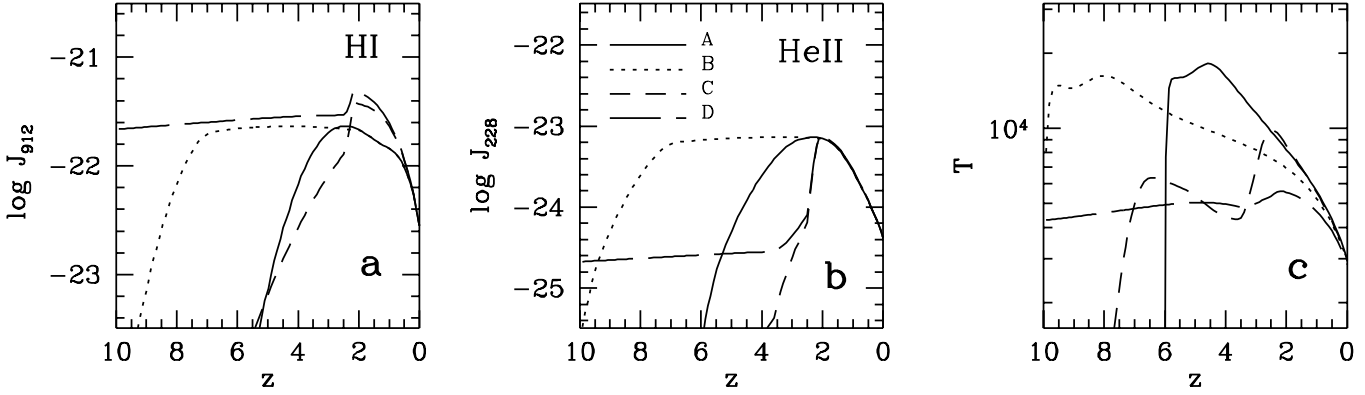,width = 18.0cm
,angle=0.}
%}
\vspace{-13.0cm}
\caption{The left and middle panels show the 
absorption cross section-weighted specific intensity 
of the UV background at the \hi\ and \heii\  ionization edges 
($\int{I_{\nu}\sigma\frac{\dd\nu}{\nu}}/\int{\sigma
\frac{\dd\nu}{\nu}}$)  in units
of $\erg\cm^{-2}\s^{-1}\Hz^{-1}\sr^{-1}$
for four different models. The right
panel shows the corresponding thermal history of the IGM at the mean
density. For a  description of models A--D see
section 2.\label{fig_1}}
\end{figure*}
%\vfill\newpage

\section[]{Reionization and the thermal history of a photoionized 
intergalactic medium}

We have calculated the thermal history of an IGM 
of primordial composition within a cosmological 
context using SPH simulations performed with GRAPESPH (see Steinmetz 1996
for a detailed description of the numerical techniques 
and Rauch, Haehnelt \& Steinmetz 1997 for a description of their 
absorption properties). The code includes the relevant cooling 
and heating processes and follows self-consistently the 
non-equilibrium evolution 
of the  baryonic species (H, H$^{+}$, He, He$^{+}$, He$^{++}$, and $e^-$). 
For the low-density regime  
relevant in this paper the thermal evolution of 
the IGM is rather  simple  (Miralda-Escud\'e \& Rees 1994, 
Hui \& Gnedin 1997, Steinmetz \& Haehnelt 1997). The dominant  
heat input is due to ionization of hydrogen and helium 
while the main cooling processes are Compton cooling at $z\ga 5$ 
and adiabatic  cooling at lower redshift. At reionization the gas
acquires an initial temperature 

\begin{eqnarray}
T_{i} &=& \frac{1}{3k_{\rm B}}      
\int 4\pi\, I_{\nu}\; 
\left  [(h\nu-h\nu_{\hi})\;\sigma_{\hi} \right .\nonumber \\ 
&+& \left . (h\nu-h\nu_{\hei})\;\sigma_{\hei}+ 
(h\nu-h\nu_{\heii})\;\sigma_{\heii} \right ]\; 
\frac{\dd \nu}{h\nu} \nonumber \\
& /&  
\int {4\pi\,I_{\nu}\; 
\left [\sigma_{\hi} + \sigma_{\hei}+ \sigma_{\heii} \right ]\; 
\frac{\dd \nu}{h\nu} }, \nonumber \\
\end{eqnarray}

where $\nu_{\rm X}$ and $\sigma_{\rm X}$ are  ionization energies 
and ionization absorption cross sections and $I_{\nu}$ is the specific 
intensity of the ionizing background. 
Afterwards the gas cools adiabatically within a Hubble time 
to the (redshift dependent) equilibrium temperature where 
photo-heating balances adiabatic cooling. 
The temperature of the gas therefore depends on the 
spectrum of the ionizing source and on the redshift
when reionization occurs.

\begin{figure*}
\vskip -1.75cm
%\centerline{
%for/autovp/fig103.ps,width = 18.0cm
\hskip-27.0cm \psfig{file=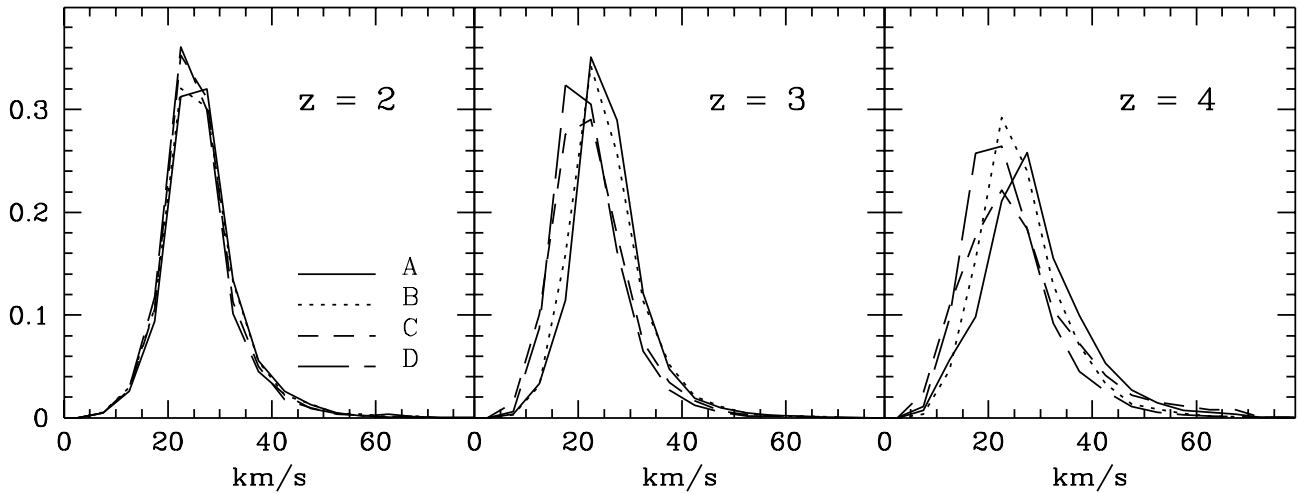,width = 18.0cm
,angle=0.}
%}
\vspace{-11.50cm}
\caption{The distribution of the Doppler parameter obtained by profile 
fitting artificial spectra for models A--D for 
$z=2,3,4$. The lines have column densities in the range 
$12.8\le N_{\hi} \le 16.0$. \label{fig_2}}
\end{figure*}
%\vfill\newpage

We have investigated four different models to cover  the 
uncertainty concerning the respective role of quasars 
and stars in the reionization of the Universe: 

\begin{list}{$\bullet$}{\setlength{\leftmargin}{17mm}
\setlength{\labelwidth}{17mm}}

\item[Model A:]
Ionizing background as proposed by Haardt 
\& Madau (1996). The spectrum is a power-law with 
$\alpha$=1.5  processed by the intervening
\op absorption. 

\item[Model B:]
Same as model A but the redshift evolution is stretched 
towards higher redshift (``Early QSO's''). 

\item[Model C:]
Same as model A but the redshift evolution is compressed  
towards lower redshift (``Late QSO's''). A stellar component 
is added to the UV background which reionizes hydrogen at redshift 
$z\sim 6$   with a soft spectrum (power law with $\alpha$=5).

\item[Model D:]
Same as C but the stellar component  reionizes hydrogen at redshift 
$z\sim 30$.   

\end{list}

Figures 1a and 1b show the evolution of the UV background for the 
four scenarios
at the \hi\ and \heii\  ionization edges respectively. 
At  redshift three the  \hi\ ionizing flux is 
1 to 3~$\times 10^{-22}\erg\cm^{-2}\s^{-1}\Hz^{-1}$, 
slightly  lower than  estimates  from the proximity effect 
but consistent with the mean flux decrement in QSO absorption spectra  
(Bechthold 1994; Giallongo et al. 1996; Cooke, Espey 
\& Carswell 1997; Rauch et al. 1997). 
Model A was chosen by Haardt \& Madau to represent the 
UV background due to observed quasars. Model B mimics the existence of
an as yet undetected population of quasar at redshifts beyond five.
Models C and D address the possibility that the Haardt \& Madau model
overestimates the UV background due to quasars at redshifts larger
than three. In the latter case the UV background would have to be 
dominated by a stellar  contribution at these redshifts.
Figure 1c shows the  temperature evolution of the 
IGM for the mean density at a given redshift. 
There are striking differences in the thermal history 
between the  models. In models  A and B  
hydrogen and helium are reionized almost simultaneously with a rather hard 
AGN-like spectrum. This leads to an initial temperature of
about $2\times 10^4\K$. Thereafter  the temperature approaches the 
redshift dependent equilibrium temperature. 
In scenario C and D the stellar component of the UV background 
is too soft to reionize helium at the same time as hydrogen. This results in
an initial temperature of $7\times 10^{3} \K$ immediately after
reionization --  significantly smaller than in model 
A and B. In Model C the late reionization of helium due to the QSO
component of the UV background leads to a corresponding jump in the 
temperature while in model D Helium reionization occurs gradually 
before the QSO component of the UV background dominates. 
In the observationally relevant redshift range 
between five and three  the difference in temperature 
between the different models is  as large as a factor 3 to 4.

\section[]{Simulated and observed Doppler parameter}  

In Figure 2 we show the  distribution of 
the Doppler parameter for the four scenarios at 
$z$=2,3,4. The Doppler parameters were obtained 
by fitting Voigt profiles to artificial spectra with the automatic line 
fitting program AUTOVP kindly provided by Romeel Dav\'e
(Dav\'e et al.~1997). 
At redshift two  the temperature in all four models is very similar 
and this is reflected by a nearly identical  distribution 
of the Doppler parameters. At $z\ge 3$ the considerable differences in the 
temperature between the four models clearly carry over to the
Doppler parameter distribution. In models  C and D 
the distribution extends to smaller values of the Doppler parameter  
than in models A and B by about $5 \kms$.

In the following we will present a preliminary  comparison between
observed and simulated Doppler parameter. It should, however,  be kept
in mind that that the primary aim of our paper is to demonstrate that 
there are observable differences between  different reionization histories.
There are a number of uncertainties in any comparison with the observed
data. The observed Doppler parameter distributions  we discuss
have been obtained with different fit procedures. The spectrum
of the UV background component  due to  star formation 
could be somewhat harder than the power law with 
$\alpha =5$ which we have assumed.  Furthermore our simulations  
do not include  any feedback effects due to star formation and may therefore 
underestimate the fraction of the IGM  in a hot collisionally ionized phase.

Figure 3 summarizes the  comparison of observed 
and simulated Doppler parameter distribution at three different 
redshifts (Lu et al 1996, Kim et al 1997).
The mean redshift and the column density range of the sample 
of lines identified with AUTOVP in the artificial spectra has been 
matched to those of the observed samples. 
At high redshift ($\expec{z}$=3.3,3.7) there is good agreement between 
the observed and simulated Doppler parameter distribution 
in the case of  models A and B. The only discrepancy is 
a modest high-velocity tail which is present in the observed data 
but not in the simulated distribution.   
Models C and D clearly disagree with the observed distribution.
They show  a considerable fraction of lines which
are cooler than the lower cut-off of 15 $\kms$. At redshift 2.3 there
is  a significant discrepancy 
between  simulated and observed distribution for all four models. 
The observed distribution shows a pronounced high-velocity tail  
which is not reproduced by any of the models. The existence of this 
high-velocity tail has been  known for some time (e.g. Carswell et al.  1987) 
and  appears to be  real (Rauch et al. 1992, Press \& Rybicki 1993).

\begin{figure*}
\vskip -1.5cm
%for/autovp/fig104.ps
\hskip-27.0cm\psfig{file=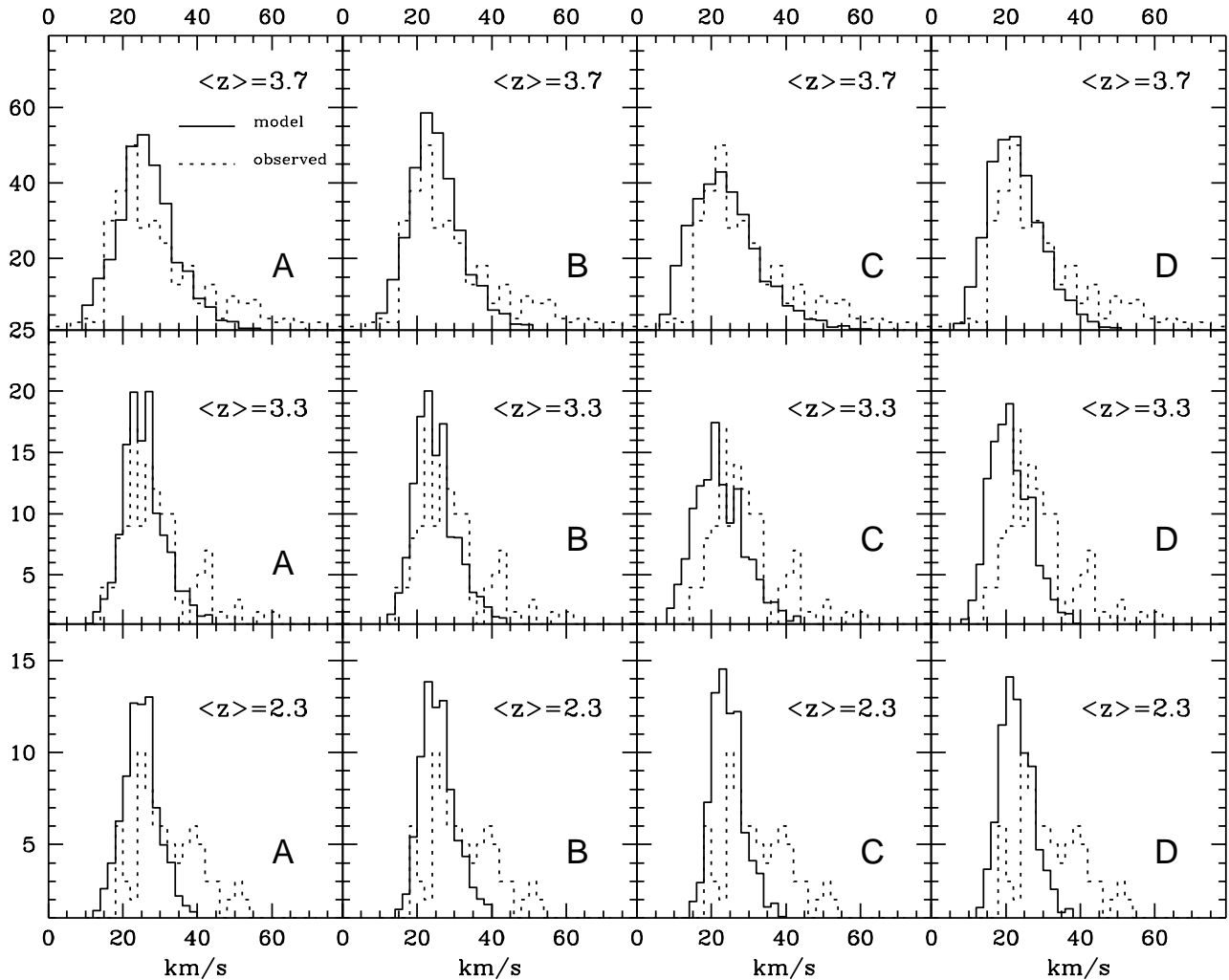,width = 18.0cm
,angle=0.}

\vspace{-6.0cm}
\caption{Observed (dashed line) and simulated (solid line)
Doppler parameter distributions for models A--D (from left to right). 
The mean redshift of the samples is $\expec{z}$=2.3,3.3,3.7.
The observed data are taken from Lu et al. (1997) and Kim et al. (1997).
The histograms of the simulated  Doppler parameter are renormalized in
each panel such that the area under the curves is the same
as that of the observed distribution. 
\label{fig_3}}
\end{figure*}
%\vfill\newpage

\section[]{Discussion and Conclusions}

We have demonstrated that the range of plausible reionization and thermal 
histories for the universe result in observable differences in the 
Doppler parameter distribution of intermediate column density 
\op absorption lines at $z \sim 3$ to $5$. 

The observed Doppler parameter distribution at $z\ga 3$ 
and the drop of the lower cut-off of the distribution from 
20 to 15 $\kms$ between redshift 2.3 and 3.7 is well reproduced 
if the universe is reionized with an AGN-like spectrum where  hydrogen
and helium are reionized simultaneously. 

In the models where  reionization occurs 
with a soft stellar spectrum  roughly  
10\% of the lines at $z \sim 3.5$ have Doppler parameters between 
10  and 15 $\kms$, well below the cut-off in the observed Doppler 
parameter distribution at this redshift (15 $\kms$).
If  the universe were to be reionized by a stellar component of the UV
background, either  its spectrum would  have to be hard enough to 
ionize helium and hydrogen at the same time,  or the QSO contribution 
would have to become dominant before redshift four.  

None of our models can explain the observed high-velocity tail 
of the observed low-redshift sample ($\expec{z}$=2.3). 
About 15 \% of all observed lines at this redshift have Doppler 
parameters of 40 to 60 $\kms$ which would correspond to 
temperatures of $1$ to $2 \times 10^5\K$ if the broadening were
purely thermal. This is well above what can plausibly  be
reached by photoionization. This may   
indicate either,  (i) that our simulations significantly underestimate the
fraction of the IGM in a hot collisionally ionized phase because they
do not include any feedback effects due to star formation
or (ii) that there is a turbulent component  to the broadening 
on scales below the resolution limit of the simulations which is more 
important at lower redshift.  

Our results suggest that a larger  and homogenous sample of
lines, extending over a wider redshift range, will be an
excellent tool to further constrain the reionization history.

\noindent
\section*{Acknowledgments}
We thank Simon White for a careful reading of the manuscript and 
are grateful to Romeel Dav\'e for providing the line-fitting 
software AUTOVP. Support by NATO grant CRG 950752 and 
the ``Sonderforschungsbereich 375-95 f\"ur Astro-Teilchenphysik der 
Deutschen  Forschungsgemeinschaft'' is also gratefully acknowledged.

\label{lastpage}


\begin{thebibliography}{}

\bibitem{}Bahcall J.N., Salpeter E.E 1965, ApJ ,142, 1677

\bibitem{}Bechthold, J. 1994, ApJS, 91,1 

\bibitem{}Bi H.G., B\"orner G., Chu Y.  1992, A\&A, 266, 1 

\bibitem{}Carswell R.F., Webb J.K., Baldwin J.A., Atwood B.,  
1987, ApJ, 319, 709 

\bibitem{}Cen R., Miralda-Escud\'e J., Ostriker J.P., Rauch M.,
1994, ApJ,  437, L9

\bibitem{}Cooke A.J., Espey B., Carswell R.F., 
1997, MNRAS, in press

\bibitem{}Dav\'e R., Hernquist L.,  Weinberg D.H., Katz N.,
1997, ApJ, in press, astro-ph/9609115

\bibitem{}Giallongo E., Cristiani S., D'Oderico S., Fontana A.,
Savaglio S., 1996, ApJ, 466, 46 

\bibitem{}Gunn J.E., Peterson B.A., 1965, ApJ, 142, 1633


\bibitem{}Haardt F., Madau P., 1996, ApJ, 461, 20

\bibitem{}Hernquist L., Katz N., Weinberg D.H.,  Miralda-Escud\'e J.,
1996, ApJ, 457, L51

\bibitem{}Hui L., Gnedin N.Y., 1997, MNRAS, in press

\bibitem{}Kim T.S., Hu E.M., Cowie L.L., Songaila A., 1997, AJ, in
press, astro-ph/9704184



\bibitem{}Lu L., Sargent W.L.W., Womble D.S., Masahide T.-H., 1996, 472, 509 

\bibitem{}Meiksin A., 1994, ApJ, 431, 109 

\bibitem{}Miralda-Escud\'e, J., Cen, R., Ostriker, J.P., Rauch, M.,
1996, ApJ, 471, 582

\bibitem{}Miralda-Escud\'e J., Rees M.J., 1994, MNRAS, 266, 343

\bibitem{}Petitjean, P., M\"ucket,  J.P., Kates, R.E., 1995, A\&A, 295, L9

\bibitem{}Press W.H., Rybicki G.B., 1993, ApJ, 418, 585 

\bibitem{}Rauch M., Haehnelt M.G., 1995, MNRAS, 275, L76 

\bibitem{}Rauch M., Haehnelt M.G.,  Steinmetz M., 1997, ApJ, 481, 601

\bibitem{}Rauch M., Miralda-Escud\'e J., Sargent W.L.W., Barlow
T.A., Weinberg D.H., Hernquist H., Katz N., Cen R., Ostriker J.P.
1997, ApJ, submitted, astro-ph/9612245

\bibitem{}Rauch M., Carswell R.G., Chaffee F.H., Foltz C.B., Webb
J.K., Weyman R.J. Bechthold J. Green R.G., 1992, ApJ, 390, 387 

\bibitem{}Steinmetz M., 1996, MNRAS, 278, 1005

\bibitem{}Steinmetz M., Haehnelt M., 1997, in preparation

\bibitem{}Weinberg D.H,  Miralda-Escud\'e J., Hernquist L., Katz,
N.,1997, ApJ, submitted, astro-ph/9701012

\bibitem{}Zhang, Y., Anninos, P., Norman, M.L., 1995, ApJ, 453, L57


\end{thebibliography}
\end{document}